\documentclass[twocolumn,showpacs,preprintnumbers,a4paper,final,tightenlines,10pt]{revtex4}
\usepackage{graphicx}
\usepackage{amsmath}

\begin{document}

\title{Decoherence dynamics of coherent electronic excited states in the
photosynthetic purple bacterium Rhodobacter sphaeroides}
\author{Xian-Ting Liang$^{1,2}$\footnote{Email: xtliang@ustc.edu},
Wei-Min Zhang$^{2,3}$\footnote{Email: wzhang@mail.ncku.edu.tw}, and
Yi-Zhong Zhuo$^{4}$} \affiliation{$^{1}$Department of Physics and
Institute of Modern Physics, Ningbo
University, Ningbo, 315211, China\\
$^{2}$Department of Physics and Center for Quantum Information Science,
National Cheng Kung University, Tainan 70101, Taiwan \\
$^{3}$National Center for Theoretical Science, Tainan 70101, Taiwan\\
$^{4}$China Institute of Atomic Energy, P.O. Box 275, Beijing 102413, China}

\begin{abstract}
In this paper, we present a theoretical description to the quantum
coherence and decoherence phenomena of energy transfer in
photosynthesis observed in a recent experiment [see Science 316,
1462 (2007)]. As a successive two-color laser pulses with selected
frequencies cast on a sample of the photosynthetic purple
bacterium Rb. sphaeroides two resonant excitations of electrons in
chromophores can be generated. However, this effective two-level
subsystem will interact with its protein environment and
decoherence is inevitable. We describe this subsystem coupled with
its environment as a dynamical spin-boson model. The non-Markovian
decoherence dynamics is described using a quasi-adiabatic
propagator path integral (QUAPI) approach. With the photon-induced
effective time-dependent level splitting energy and level flip coupling
coefficient between the two excited states and the environment-induced
non-Markovian decoherence dynamics, our theoretical result is
in good agreement with the experimental data.

\end{abstract}

\maketitle

\section{Introduction}  In the past
decades, increasing attention has been paid to understand the
intrinsic mechanisms of the efficient energy transfer in light
harvesting complexes. In the previous
investigations, the energy transfer was often described by a
semiclassical model that invokes `hopping' excited-state
populations along discrete energy levels. Explicitly, the
electrons in pigment molecules, chromophores, are excited by the
incident light. Then the electronic excitation moves downhill from
energy level to energy level through the chromophores before being
trapped in the reaction center. However, it has not been conceived
that the high efficient energy transfers between chromophores is
realized through the electronic excited state hopping.

To understand the energy transfer among excited states of each
chromophore in the Rhodobacter (Rb.) sphaeroides reaction centers, Fleming's group
proposed recently that the plants and the photosynthetic bacteria
may utilize a `clever' quantum algorithm for the energy transfer
\cite{Science_316_1462, Nature_446_2007_782}. Unlike the
semiclassical model, Fleming and co-workers suggested that a
successive excitations make a coherence superposition between
electronic excited states of the chromophores. The energy transfer
is considered to be carried by the coherence of the superposition
state through an easiest route analogous to Grover's algorithm for
the fastest possible search of an unsorted information database
\cite{PhysRevLett79_325_1997}. However, Mohsehi \textit{et al}.
\cite{qu08064725} argued that the standard Grover's algorithm may
not be applicable to the energy transfer mechanism in
photosynthesis. Nevertheless, even if the plants and bacteria work
unlike the Grover's algorithm, the coherent dynamics evolution of
the electronic excited states in chromophores could still play an
important role in the chromophoric energy transfer
\cite{NewJPhys10_113019_2008,PhysRevLett96_028103_2006}. In this
paper we shall derive a dynamical spin-boson model to describe how
long the coherent superposition state can be persisted in the
photosynthetic purple bacterium Rb.~sphaeroides.

\section{Dynamical Spin-Boson Model} Recent experiments by Fleming
group showed that the electronic coherence between two excited
states may last for $400$ fs or longer \cite{Science_316_1462,
Nature_446_2007_782}. Instead of investigating the coherence among
all excited states of each chromophores in the Rb.~sphaeroides reaction centers
\cite{PhysRevLett_92_2004_218301,PhysRevLett77_4675_1996}, they
investigated the coherent superposition state of two electronic
excited states $\left\vert H\right\rangle $ and $\left\vert
B\right\rangle $ in chromophores created experimentally from two
successive laser pulses. The laser pulses (with wavelength $750$
nm and $800$ nm) derives the system which is initially in the
ground state $|g\rangle $ into a coherent superposition state
$|\Psi \rangle =\alpha |H\rangle +\beta |B\rangle.$
However, the system of the two excited states inevitably interacts
with its environment, which results in the decoherence of the
coherent superposition state $|\Psi \rangle $. To obtain the persistent time of
coherence, i.e., the decoherence time of $|\Psi \rangle $, Fleming
et al. use the third incident pulse (with wavelength $750$ nm) to
produce a photon echo. From the echo signals they extract the
decoherence time of the coherent state $|\Psi \rangle $
\cite{PhysRevLett100_013603_2008}.

To be specific, in the two-color photon echo experiment on bacterial
reaction centers (RC) \cite{Science_316_1462}, the RC from the
photosynthetic purple bacterium Rb.~sphaeroides includes a
bacteriochlorophyll dimer, called the special pair (p), in the
center, an accessory bacteriochlorophyll flanking p on each side
(BChl), and a bacteriopheophytin (BPhy) next to each BChl. The RC of
Rb. sphaeroides has several absorption peaks which are made by the
chromophores. In Ref.~\cite{Science_316_1462}, the absorption
spectrum of the p-oxidized RC at $77$ K shows the H band at $750$ nm
and the B band at $800$ nm (where H and B are used to denote
excitonic states which are dominatingly produced from monomeric BPhy
and accessory BChl in the RC, respectively). At the first step, they
used a successive two laser pulses with different colors and tuned
for resonant excitation of H transition at $750$ nm (at time $\tau
_{1}^{\prime })$ and the B transition at $800$ nm (at time $\tau
_{2}^{\prime })$ cast on the sample. The two pulses produce a
coherent superposition state $|\Psi \rangle$ between the electronic
excited states $\left\vert H\right\rangle $ and $\left\vert
B\right\rangle$. Then, in order to measure the decoherence time of
the coherent superposition of these two excited states, the third
laser pulse is cast on the sample after a time $t_{2}$ from the
second pulse, which generates a photon echo. When the times
$t_{1}=\tau _{2}^{\prime }-\tau _{1}^{\prime }$ and $t_{2}=\tau
_{3}^{\prime }-\tau _{2}^{\prime }$ are different one can detect the
different integrated intensity of the echo signals in the phase
matched direction. If the time $t_{1}$ is fixed (the fixed time
$t_{1}=30$ fs in Ref. \cite{Science_316_1462} ), the integrated echo
signals as a function of $t_{2}$ represents the decoherence of the
coherent superposition between $\left\vert H\right\rangle $ and
$\left\vert B\right\rangle $.

The evolutions of the integrated echo signals are plotted in the
Fig.~3 of Ref.~\cite{Science_316_1462}. From the experimental data
one can see that: (1) The coherence is resonant with different
frequencies in the first $400$ fs. The first, the second and the
third periods are about $100$ fs, $120$ fs, and $130$ fs
respectively; (2) The coherence between the excited states
$\left\vert H\right\rangle $ and $\left\vert B\right\rangle $
persists for more than $400$ fs; (3) The third peak in the evolution
of the coherence decay more fast than the first and second peaks. In
the following, we shall attempt to understand theoretically the
dynamical process of excited electron states in chromophores and to
explain the above experimental result. It is interesting to see, as
we shall show later, that the problem can be described by a
dynamical spin-boson model where the two excited states are
time-dependently coupled one another through the laser pulses, as an
effect of the photon-induced dynamics. To be explicit, we model the
BChl and BPhy molecules by the Hamiltonian in the Condon
approximation as
\cite{Mukamel-book,JChemPhys107_3876_1997,JChemPhys124_234504_2006}
\begin{align}
H_{e}=& \epsilon _{0}\left\vert g\right\rangle \left\langle g\right\vert
+\epsilon _{H}\left\vert H\right\rangle \left\langle H\right\vert +\epsilon
_{B}\left\vert B\right\rangle \left\langle B\right\vert  \notag \\
& +\epsilon _{HB}|HB\rangle \langle HB|+J_{0}\left( \left\vert
H\right\rangle \left\langle B\right\vert +\left\vert B\right\rangle
\left\langle H\right\vert \right)  \notag \\
& +\sum_{j=H,B}\vec{\mu}_{j}\cdot \vec{E}\left( t\right) \big(\left\vert
g\right\rangle \left\langle j\right\vert +\left\vert j\right\rangle
\left\langle g\right\vert \big).  \label{bareh}
\end{align}
Here, $\epsilon _{0}$, $\epsilon _{H}$, $\epsilon _{B}$ and
$\epsilon _{HB}$ are the energies of the ground state $|g\rangle $,
the two excited states $\left\vert H\right\rangle $ and $\left\vert
B\right\rangle $, and the doubly excited state $|HB\rangle $.
$J_{0}$ is the electronic coupling between the two excited states.
$\vec{\mu}_{j}$ $\left( j=H,B\right) $ is the corresponding
electronic dipole, and $\vec{E}\left( t\right) $ is the external
electronic filed of the two successive laser pulses, namely,
$\vec{E}\left( t\right) =\vec{E}_{01}e^{-\Gamma _{1}\left(
t+t_{1}\right) ^{2}+i\omega _{H}\left( t+t_{1}\right)
}+\vec{E}_{02}e^{-\Gamma _{2} t^{2}+i\omega _{B}t } $ where $\Gamma
_{1,2}$ are the decay constants of the laser pulses as they pass
though the sample.

For simplicity, we assume that the dipole moments for the H and B
states are the same: $\vec{\mu}_{H}=\vec{\mu}_{B}\equiv
\vec{\mu}.$ We can decouple the states $|g\rangle ,$ $|HB\rangle $
from $|H\rangle ,$ $ |B\rangle $ by making the following canonical
transformation to Eq.~(\ref{bareh}):
\begin{align}
&\mathcal{H}_{e}=e^{S}H_{e}e^{-S} , \\
&S=\kappa (t)\big[\alpha \left(
\left\vert H\right\rangle \left\langle g\right\vert -\left\vert
g\right\rangle \left\langle H\right\vert \right) \notag \\
&~~~~~~+\beta
\left(\left\vert B\right\rangle \left\langle g\right\vert -
\left\vert g\right\rangle \left\langle B\right\vert \right) \big],
\end{align}
where $\kappa (t)=\vec{\mu}\cdot \vec{E }\left( t\right) $, $\alpha
=(\epsilon _{B}-\epsilon _{0}-J_{0})/\Omega $, $ \beta =(\epsilon
_{H}-\epsilon _{0}-J_{0})/\Omega $, and $\Omega =(\epsilon
_{H}-\epsilon _{0})(\epsilon _{B}-\epsilon _{0})-J_{0}^{2}$. Under
the condition $\kappa(t)\alpha \approx \kappa(t) \beta \ll 1$, we
obtain
\begin{align}
\mathcal{H}_{e} =\big\{\epsilon _{0}^{\prime } &\left\vert
g\right\rangle \left\langle g\right\vert +\epsilon _{HB}|HB\rangle
\langle HB| \big\}+ \big\{\epsilon _{H}^{\prime }\left\vert
H\right\rangle \left\langle
H\right\vert  \notag \\
& +\epsilon _{B}^{\prime }\left\vert B\right\rangle \left\langle
B\right\vert +J_{\rm eff}\left( \left\vert H\right\rangle
\left\langle B\right\vert +\left\vert B\right\rangle \left\langle
H\right\vert \right)\big\} , \label{effH1}
\end{align}
where, $\epsilon _{0}^{\prime }=\epsilon _{0}-\kappa ^{2}(t)(\alpha
+\beta )$, $\epsilon _{H}^{\prime }=\epsilon _{H}+\kappa
^{2}(t)\alpha $, $\epsilon _{B}^{\prime }=\epsilon _{B}+\kappa
^{2}(t)\beta $ and $J_{\rm eff}=J_{0}+\kappa ^{2}(t)(\alpha +\beta
)/2$. As one can see, the ground state $|g\rangle$ and the doubly excited
state $|HB\rangle$ are now decoupled from the single excited states
$|H\rangle$ and $|B\rangle$. Thus the decoherence dynamics of the
coherent superposition state $|\Psi \rangle $ is fully determined by
the effective two-level Hamiltonian, the second curly bracket in
Eq.~(\ref{effH1}) which can be rewritten as
\begin{equation}
H_0(t)=\frac{\epsilon \left( t\right) }{2}\sigma _{z}+\frac{\Delta
\left( t\right) }{2}\sigma _{x}.  \label{effH2}
\end{equation}
Here, $\sigma _{i}\left( i=x,\text{ }z\right) $ are the Pauli
matrix, $\epsilon \left( t\right) = \epsilon _{H}-\epsilon _{B}
+\kappa ^{2}(t)\left( \alpha - \beta \right) $, $\Delta \left(
t\right) =2J_{0}+\kappa ^{2}(t)(\alpha +\beta )$. It shows that
the energy splitting of the two excited states and the coupling
between them are shifted by the pulse-induced time-dependent
dipole-dipole interaction [$\sim \kappa^2(t)$], as an effect of
photon-induced dynamics.

Furthermore, this two-level system, as a part of chromophores, is
inevitable to interact with its protein environment through the
thermal vibrations. This thermal reservoir can always be modeled
with a set of harmonic oscillators. Then the total system of the
two-level system coupled to its environment can be written  as a
dynamical spin-boson model,
\begin{equation}
H=H_{0}(t)+\sigma _{z}\sum_{i}c_{i}(b_{i}^{\dagger
}+b_{i})+\sum_{i}\hbar \omega _{i}b_{i}^{\dagger }b_{i}
\label{totalH} ,
\end{equation}%
in which the last two terms are the interaction between the
two-level system and the bath and the Hamiltonian of the thermal
bath itself, respectively. The parameter $c_{i}$ is the coupling
of the system to the bath and $ b_{i}^{\dagger }$ ($b_{i})$ are
the creation (annihilation) operators of the $i$-th thermal mode.
The decoherence dynamics of the above two-level system are
mainly induced by the back-actions of the thermal bath as well as the
decay of the time-dependent coupling induced by laser pulses. The
solution of the problem is now completely determined by the
parameters $\varepsilon(t), \Delta(t)$ and the spectral density of
the thermal bath\cite{RevModPhys_59_1}
\begin{equation}
J\left( \omega \right) =\frac{\pi }{2}\sum_{i}c_{i}^{2} \delta
\left( \omega -\omega _{i}\right) =\frac{\pi }{2}\hbar \xi
_{s}\omega \Big(\frac{\omega}{\omega_c}\Big)^{s-1}e^{-\omega /\omega
_{c}}. \label{eq4}
\end{equation}%
Here $\omega _{c}$ is the high-frequency cut-off of the bath modes.
Different $s$ correspond to the super-Ohmic $(s>1)$, the Ohmic
$\left( s=1\right) $, and the sub-Ohmic $\left( 0\leq s<1\right)$
baths, and $\xi _{s}$ is a dimensionless dissipative parameter
describing the coupling between the two-level system and the bath.

\section{Numerical Method}
The spin-boson model has been
investigated by many methods though it has not be exactly solved
and some approximation, such as the Markov approximation, must be used \cite{Weiss,
RevModPhys_59_1}. Here, we shall apply the quasiadiabatic propagator
path integral (QUAPI) technique \cite{ChemPhysLett_221_482} to
explore the decoherence dynamics of this system, where
non-Markovian processes are involved.
Indeed, as it has been pointed out \cite{Science_316_1462}
the evolution of this time-dependent system should be highly
non-Markovian. To see how the non-Markovian dynamics may play an
important role, it is useful to estimate the correlation time of
the thermal bath, which can be obtained from the bath response
function
\begin{equation}
C\left( t\right) =\frac{1}{\pi }\int_{0}^{\infty }d\omega J\left( \omega
\right) \left[ \coth \left( \frac{\beta \hbar \omega }{2}\right) \cos \omega
t-i\sin \omega t\right] ,  \label{eq6}
\end{equation}%
where $\beta =1/k_{B}T$ with Boltzmann's constant $k_{B}$ and the
temperature $T$. When the real and imaginary parts of $C(t)$ behave
as a delta function $\delta \left( t\right) $ and its derivative
$\delta ^{\prime }\left( t\right) $, respectively, the dynamics of
the reduced density matrix is Markovian. Otherwise, non-Markovian
dynamics occurs. The broader the Re$[C\left( t\right) ]$ and
Im$[C\left( t\right) ]$ are, the longer the correlation time will
be, and the more serious the practical dynamics is distorted by the
Markov approximation. Similar to Ref.~\cite{PhysRevB_72_245328}, we
calculate the correlation times of the bath. The result is shown in
Fig.~1. As we can see the correlation time of the bath is about
$\tau _{c}\approx 15$ fs which actually depends on the frequency
cutoff $\omega _{c}$ in the bath.
\begin{figure}[tbp]
\begin{center}
\scalebox{0.4}{\includegraphics{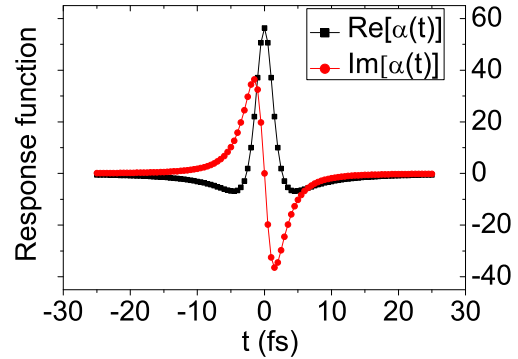}}
\end{center}
\caption{(Color online) Real and imaginary parts of the response
functions ($C(t)$) for Ohmic bath. Here, we take $\xi _{1}=0.75,$
$\hbar \omega _{c}=2000$ cm$^{-1}$ and $T=77$ K$.$} \label{fig1}
\end{figure}

The decoherence dynamics of the coherent superposition states
$|\Psi \rangle $ is characterized by the time evolution of the
reduced density matrix obtained after tracing out the bath degrees
of freedom, i.e.,
\begin{equation}
\rho \left(t\right)
=\text{Tr}_{\rm bath} \big[
e^{-iHt/\hbar }\rho_{\mathrm{tot}}\left( 0\right) e^{iHt/\hbar
}\big] .  \label{eq9}
\end{equation}
Following the experiment, the interaction between system and bath
is turned on at $t=\tau'_2$ when the second laser pulse is
applied. Thus, the density matrix of the total system before the
time $t=\tau'_2$ is a direct product of the system and bath
components, namely, $\rho _{\mathrm{tot}}\left( \tau'_2\right)
=\rho \left( \tau'_2 \right) \otimes \rho _{\rm bath}\left(
\tau'_2 \right) ,$ where $\rho \left( \tau'_2 \right) $ and $\rho
_{\rm bath}\left( \tau'_2 \right) $ are the "initial" states of
the system and the bath at $t=\tau'_2$. If we set
$\vec{E}_{01}=\vec{E}_{02}\equiv \vec{E}_{0}$, the initial
resonant two excited state can be written as $\left\vert \Psi
\right\rangle =\frac{1}{\sqrt{2}}\left( \left\vert H\right\rangle
+\left\vert B\right\rangle \right) $. After the canonical
transformation, the initial state remains almost the same:
$\left\vert \Psi ^{\prime }\right\rangle =e^{S}\left\vert \Psi
\right\rangle \approx \frac{1}{\sqrt{2}}\left( \left\vert
H\right\rangle +\left\vert B\right\rangle \right) $ under the
condition $\kappa \sqrt{\alpha ^{2}+\beta ^{2}}\ll 1,$ which is
satisfied for the parameters we taken in the following numerical
calculations. We also set the bath initially at the thermal
equilibrium, namely, $\rho _{\rm bath}\left( \tau'_2 \right)
=e^{-\beta H_{b}}/$Tr$\left( e^{-\beta H_{b}}\right) $.

The reduced density matrix $\rho (t)$ can be evaluated by using the
well established iterative tensor multiplication (ITM) algorithm
derived from the QUAPI. This algorithm is numerically exact and
successfully tested and adopted in various problems of open quantum
systems \cite{JChemPhys_102_4600,PhysRevE_62_5808}. For details of
the scheme, we refer to previous works \cite{ChemPhysLett_221_482}.
The QUAPI asks for the system Hamiltonian splitting into two parts
$H_{0}$ and $H_{env}$, where $ H_{env}=H_{e-b}+H_{b}$. In order to
make the calculations converge we use the time step $\Delta t=5$ fs
which is shorter than the correlation time of the bath and the
characteristic time of the two-level subsystem. In order to include
all non-Markovian effect of the bath in the ITM scheme, one should
choose $\Delta k_{\max }$ so that $\Delta k_{\max }\Delta t$ is not
much shorter than the correlation time $\tau _{c}$ of the bath.
Here, $\Delta k_{\max }$ is roughly equal to the number of time
steps needed to span the half-width of the response function
$C(t-t^{\prime })$ \cite{ChemPhysLett_221_482}. Then taking $\Delta
k_{\max }=3$ should be large enough in our calculations.

\section{Results and discussions} The decoherence of the two
exciton states is reflected through the decays of the off-diagonal
reduced density matrix element. In Fig.~2a and b we plot the
evolution of the off-diagonal reduced density matrix element at
temperature $T=77$ K and $T=180$ K, respectively, where the
environment is assumed as an Ohmic bath ($s=1$), and the parameters
$\epsilon _{H}=12108$ cm$^{-1},$ $\epsilon _{B}=12000$
cm$^{-1},$ $\epsilon _{0}=10570$ cm$^{-1},$ $J_{0}=20$ cm$^{-1}$,
$\kappa _{0}=\vec{\mu}\cdot \vec{E}_{0}=210$ cm$^{-1}$ and $\Gamma
_{1}=\Gamma _{2}\equiv \Gamma =3\times 10^{24}$ s$^{-2}$. We take
the Kondo parameter $\xi _{1}=0.75$ and the frequency cutoff $\hbar
\omega _{c}=2000$ cm$^{-1}$ \cite{PNAS93_3926_1996} for the thermal
bath. According to Ref. \cite{Science_316_1462}, we take $t_{1}=30$
fs, and $t_{2}=50$ fs.
\begin{figure}[tbp]
\begin{center}
\scalebox{0.4}{\includegraphics{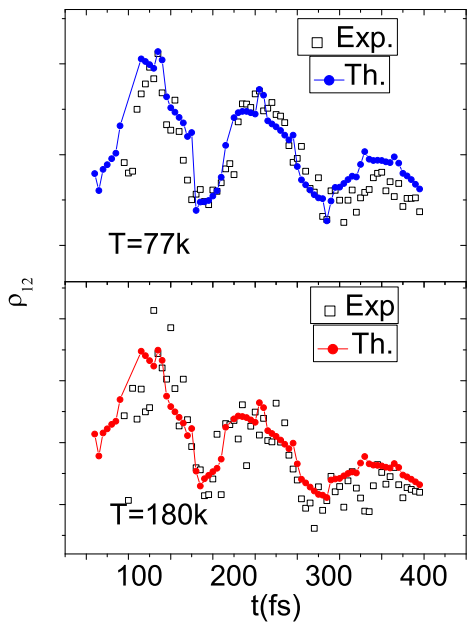}}
\end{center}
\caption{(Color online) The evolutions of the off-diagonal
coherent terms of the reduced density matrix for the two-level
subsystem with the Ohmic bath at temperature (a) $77$ K and (b)
$180$ K. } \label{fig2}
\end{figure}
The black square boxes in Fig.~2 are the experimental data from
\cite{Science_316_1462}, the linked blue and red dots are our
numerical results. The plots show that our theoretical description
is in good agreement with the experimental results. As we see the
coherence decay is much faster after the second peak and the
oscillation periods of the evolutions increase with the time. The
oscillation dephasing behaviors indeed come from the time-dependent
level splitting $\epsilon \left( t\right) $ and coupling $\Delta
\left( t\right) $ induced by laser pulses as well as the
non-Markovian processes due to the interaction between the
two-level system and the thermal bath.

\begin{figure}[tbp]
\begin{center}
\scalebox{0.4}{\includegraphics{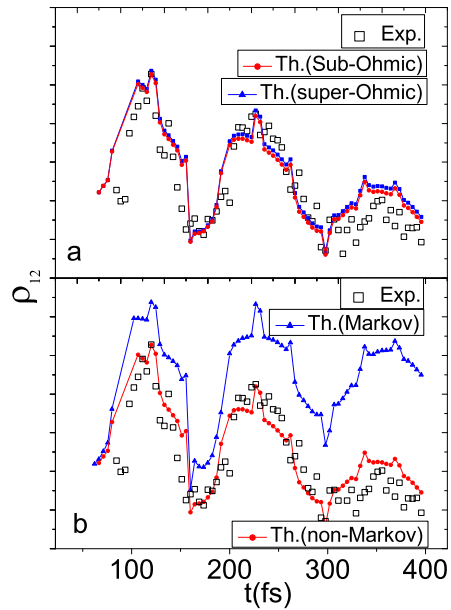}}
\end{center}
\caption{(Color online) The evolutions of the off-diagonal
coherent terms of the reduced density matrix for the two-level
subsystem in sub-Ohmic and super-Ohmic Baths. Here, we set $\xi
_{1/2}=0.07$ and $\xi_{2}=50$ at $T=77$ K. Other parameters are
the same as that in Fig. 2}\label{fig3}
\end{figure}

Further calculations show that using the sub-Ohmic bath with
$\xi_{1/2}=0.07$ and super-Ohmic bath with $ \xi _{2}=50$, we can
obtain the similar results, as show in Fig.~\ref{fig3}a. However, if
we use the same value of the dimensionless dissipative parameter
$\xi _{s}$ for the Ohmic, sub-Ohmic and super-Ohmic spectral
densities, the corresponding decoherence behaviors are completely
different. This indicates that the decoherence dynamics is
non-Markovian. To show the non-Markovian effect, we also calculate
the Markov approximation with the same parameters in the Ohmic case.
The detailed numerical results are plotted in Fig.~\ref{fig3}b,
where the difference between the non-Markovian dynamics and the
corresponding Markov approximation is obvious.

\begin{figure}[tbp]
\begin{center}
\scalebox{0.4}{\includegraphics{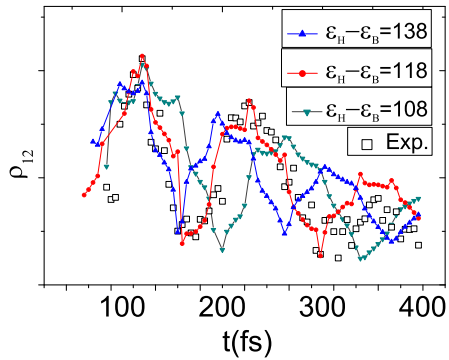}}
\end{center}
\caption{(Color online) The evolutions of the off-diagonal
coherent terms of the reduced density matrix in Ohmic bath, with
different bare level splitting $\epsilon _{H}-\epsilon _{B}=108,
118, 158$ cm$^{-1}$ at $T=77$ K. Other parameters are the same as
that in Fig. 2}\label{fig4}
\end{figure}

We should also point out that the bare level splitting $\epsilon
_{H}-\epsilon _{B} $ is not the experimentally observed values. The
experimental values ($\sim 600$ cm$^{-1}$) of the level splitting
contains the pulse-induced time-dependent effects plus the
back-action effect from the environment in the non-Markovian regime
so that the input value of $\epsilon _{H}-\epsilon _{B}$ is
different from the splitting of experimental linear absorbance
peaks. A practical calculation of the true site energy difference
that contains all the the pulse-induced time-dependent effects plus
the back-action effect from the environment may be possible
\cite{Ado062778} but it is beyond the scope of the present
investigation. Here the input value of the bare level splitting is
based on how better to fit the experimental data. Our numerical
calculations show that the profiles of the evolution curves for the
off-diagonal reduced density matrix elements are sensitive to the
changes of the bare level splitting $\epsilon _{H}-\epsilon _{B}$,
the bare coupling $J_0$ and the parameter $\kappa _{0}$ but
insensitive to the changes of $\epsilon _{H}$ and $\epsilon _{B}$
with $\epsilon _{H}-\epsilon _{B}\approx 110$ cm$^{-1}$. In
Fig.~\ref{fig4}, we plot the time evolution of off-diagonal reduced
density matrix element with different $\epsilon _{H}-\epsilon _{B}$.
As we can see, the change of the bare level splitting $\epsilon
_{H}-\epsilon _{B}$ produces a quite different pattern of the
decoherence dynamics. A smaller value of $\epsilon _{H}-\epsilon
_{B}$ will result in a shorter oscillation period of the evolution
in time.

Also, the bare electronic coupling $J_0$ between the exciton states
$|B\rangle$ and $|H\rangle$ is not the electronic coupling $J$
between the BChl and BPhy molecules that used in
\cite{Science_316_1462}. Our numerical results suggest that $J_0$ is
much smaller ($< 30$ cm$^{-1}$) than the pulse-induced dipole-dipole
interaction but is not negligible. Change of $J_0$ leads to very
different decoherence behavior as shown in Fig.~\ref{fig5}, which is
actually also convinced in the supplemental material of
\cite{Science_316_1462}. The resulting effective electronic coupling
$J_{\rm eff}$ oscillates in time with the maximum amplitude being
less than $400$ cm$^{-1}.$
\begin{figure}[tbp]
\begin{center}
\scalebox{0.4}{\includegraphics{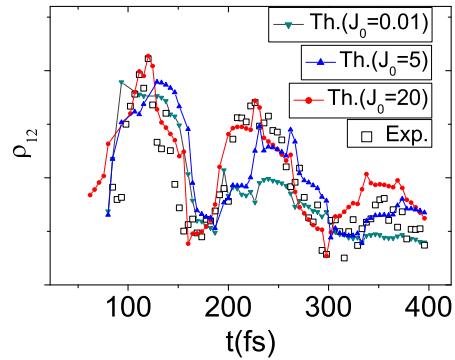}}
\end{center}
\caption{(Color online) The evolutions of the off-diagonal
coherent terms of the reduced density matrix for the two-level
subsystem in Ohmic bath at $T=77$ K, with different bare
electronic coupling $J_{0}=20, 5$, and $0.01$ cm$^{-1}$. Other
parameters are the same as that in Fig.~2.} \label{fig5}
\end{figure}
As one can see, our numerical result shows that the oscillation of
the off-diagonal reduced density matrix element is due to the
effective time-dependence of the level splitting and flip coupling
between the two excitonic states induced by the laser pulses. This
differs from the interpretation given in \cite{Science_316_1462}
where it has been pointed out that reproducing the long dephasing
time observed would require a value of $J>320$ cm$^{-1}$, but a
large $J$ implies that the BChl and BPhy excitations are almost
degenerate, which may not be consistent with the observed spectra.
Therefore they argued that the vibrational modes begin to contribute
in the energy transfer. Their simulation shows that a vibrational
mode (with a frequency $\omega=250$ cm$^{-1}$) coupling to the
excitonic states can fit the long dephasing time very well. Here we
show that the large effective coupling between the two excited
states induced by the laser pulses combining with a very smaller but
not negligible bare coupling $J_0$ can reproduce the long dephasing
time observed, without including the contribution arisen from the
vibrational mode.

\begin{figure}[tbp]
\begin{center}
\scalebox{0.4}{\includegraphics{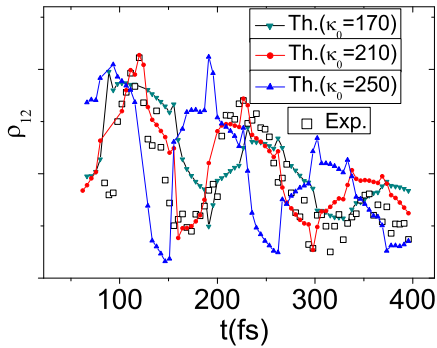}}
\end{center}
\caption{(Color online) The evolutions of the off-diagonal coherent
terms of the reduced density matrix for the two-level subsystem with
different pulse field strength, $\kappa_{0} =\vec{\mu} \cdot
\vec{E}_0=170, 210$ and $250$ cm$^{-1}$, in Ohmic bath at $T=77$ K .
Other parameters are the same as that in Fig. 2} \label{fig6}
\end{figure}

On the other hand, the level splitting and flip coupling we derived
explicitly depend on the parameter $\kappa _{0}=\vec{\mu}\cdot
\vec{E}_{0}$. The dipole moment $\mu=\sqrt{D}$ (debye) where the
dipole strength $D$ can be estimated by: $D \simeq 0.0196 n
\epsilon_{\rm max} \delta/\lambda_{\rm max}$ \cite{Kno03497}. Here
$n=1.359$ is the refractive index, $\epsilon_{\rm max}$ and $\delta$
are the half-width and the peak value at $\lambda_{\rm max}=750$ nm
and $800$ nm for H and B excited states, respectively. Fig.~1 of
\cite{Science_316_1462} shows that $\epsilon^{\rm
H}_{max}=0.5\epsilon^{\rm B}_{max}$ and $\delta_{\rm P}\simeq 1.1
\delta_{\rm B}$. Taking $D_{\rm B} =40$ debye$^2$ (corresponding to
$\mu=6.3$ debye for 800 nm BChl \cite{Sch8621}), we have $D_{\rm H}
\simeq 23.5$ debye$^2$. Averaging the dipole strength
$D=(40+23.5)/2$ for BChl and BPhy, we obtain $\mu \simeq 5.63$ debye
$=1.88 \times 10^{-27}$ C cm.  The power intensity used in Fleming's
experiment \cite{Science_316_1462} is $P=1.3 \times 10^{-4}$
J/cm$^2$ while the pulse duration $\Delta t=40$ fs. Therefore the
pulse intensity is $I_p=P/\Delta t =3.25 \times 10^9$ W/cm$^2$ which
corresponds to a pulse field amplitude $E_0=\sqrt{2}E_{\rm rm}\simeq
2.21 \times 10^6$ V/cm. Thus, the realistic dipole-field coupling
strength $\kappa _{0}=\vec{\mu} \cdot \vec{E}_{0}\simeq 209 $
cm$^{-1}$. This is very close to the value of our theoretical best
fitting $\kappa_0=210$ cm$^{-1}$ in the numerical calculation.  We
also find that the laser pulse strength $E_{0}$ controls the decay
curves. In Fig.~\ref{fig6}, we plot the evolution of the
off-diagonal reduced density matrix element to demonstrate the
$\kappa_0$-dependence of the dephasing time.
 It shows that increasing $E_{0}$
will decrease the oscillation periods of the evolution in time, and
vice versa. This property may be used for further experimental test
of whether the pulse-induced time-dependent dipole-dipole
interaction or the additional vibrational mode gives rise to the
oscillation decay of the coherent excitonic states.

In conclusion, we present in this paper a dynamical spin-boson model
with the effective time-dependence of the level splitting and flip
coupling between the two excited states, induced by laser pulses, to
describe the long dephasing time observed recently in a
photosynthesis experiment \cite{Science_316_1462}. We use the
quasi-adiabatic propagator path integral approach to analyze in
detail the non-Markovian decoherence behaviors of the two excitonic
states, $|H\rangle$ and $|B\rangle$. Due to the photon-induced
time-dependence of the energy level splitting and flip coupling
coefficient between the two excited states and the
environment-induced non-Markovian dynamics, our theoretical result
is in good agreement with the experimental data.

\begin{acknowledgments}
We would like to thank Prof. Fleming for sending us their experimental data
for comparisons. This project was sponsored by National Natural Science
Foundation of China (Grant No. 10675066), the National Science Council of
ROC under Contract No. NSC-96-2112-M-006-011-MY3, and K.C.Wong Magna
Foundation in Ningbo University.
\end{acknowledgments}


\begin{thebibliography}{99}
\bibitem{Science_316_1462} H. Lee, Y.-C. Cheng, and G. R. Fleming, Science
316, 1462 (2007).

\bibitem{Nature_446_2007_782} G. S. Engel, T. R. Calhoun, E. L. Read, T. -K.
Ahn, T. Man\v{c}al, Y. -C. Cheng, R. E. Blankenship, and G. R. Fleming,
Nature 446, 782 (2007).

\bibitem{PhysRevLett79_325_1997} L. K. Grover, Phys. Rev. Lett. 79, 325
(1997).

\bibitem{qu08064725} M. Mohseni, P. Rebentrost, and A. Aspuru-Guzik, J.
Chem. Phys. 129, 174106 (2008); P. Rebentrost, M. Mohseni, and A.
Aspuru-Guzik, arXiv: 0806.4725v1;

\bibitem{NewJPhys10_113019_2008} M. B. Plenio, S. F. Huelga, New J. Phys.
10, 113019 (2008).

\bibitem{PhysRevLett96_028103_2006} Y. C. Cheng and R. J. Silbey, Phys. Rev.
Lett. 96, 028103 (2006).

\bibitem{PhysRevLett_92_2004_218301} S. Jang, M. D. Newton, and R. J.
Silbey, Phys. Rev. Lett. 92, 218301 (2004).

\bibitem{PhysRevLett77_4675_1996} D. Leupold, H. Stiel, K. Teuchner, F.
Nowak, W. Sandner, B. \"{U}cker, and H. Scheer, Phys. Rev. Lett. 77, 4675
(1996).

\bibitem{PhysRevLett100_013603_2008} V. O. Lorenz, S. Mukamel, W. Zhuang,
and S. T. Cundiff, Phys. Rev. Lett. 100, 013603 (2008).

\bibitem{Weiss} U. Weiss, \emph{Quantum Dissipative Systems}, 2nd ed.,
(World Scientific Publishing, Singapore, 1999).

\bibitem{RevModPhys_59_1} A. J. Leggett, S. Chakravarty, A. T. Dorsey, M. P.
A. Fisher, A. Garg, and W. Zwerger, Rev. Mod. Phys. 59, 1 (1987).

\bibitem{ChemPhysLett_221_482} D. E. Makarov and N. Makri, Chem. Phys. Lett.
221 (1994) 482.

\bibitem{Mukamel-book} S. Mukamel, \emph{Principles of Nonlinear Optics and
Spectroscopy}, New York: Oxford University Press (1995).

\bibitem{JChemPhys107_3876_1997} T. Meier, Y. Zhao, V. Chernyak, and S.
Mukamel, J. Chem. Phys. 107, 3876 (1997).

\bibitem{JChemPhys124_234504_2006} T. Man\v{c}al, A. V. Pisliakov, and G. R.
Fleming, J. Chem. Phys. 124, 234504 (2006); A. V. Pisliakov, T.
Man\v{c}al, and G. R. Fleming, \textit{ibid.} 124, 234505 (2006).

\bibitem{JChemPhys_102_4600} N. Makri, and D. E. Makarov, J. Chem.
Phys. 102, 4600; 102, 4611 (1995).

\bibitem{PhysRevB_72_245328} X. -T. Liang, Phys. Rev. B 72 (2005) 245328.

\bibitem{PhysRevE_62_5808} M. Thorwart, P. Reimann, and P. H\"{a}nggi, Phys.
Rev. E 62, 5808 ( 2000).

\bibitem{PNAS93_3926_1996} N. Makri, E. Sim, D. E. Makarov, and M. Topaler,
Proc. Natl. Acad. Sci. USA 93, 3926 (1996).

\bibitem{Ado062778} J. Adolphs and T. Renger, Biophys. J. \textbf{91}, 2778 (2006).

\bibitem{Kno03497} R. S. Knox and B. Q. Spring, Photochem. Photobio.,
 77(5), 497 (2003).

\bibitem{Sch8621} A. Scherz and W. Parson, Photosynthesis Research, 9, 21 (1986).
\end{thebibliography}
\end{document}